\RequirePackage[left]{lineno}
\documentclass[aps,prl,twocolumn,superscriptaddress,amsmath,amssymb]{revtex4}
\usepackage{graphicx} 
\usepackage{epstopdf} 
\usepackage{dcolumn} 
\usepackage{xcolor} 
\usepackage{amsmath,amssymb} 
\usepackage{hyperref} 
\usepackage[utf8]{inputenc} 
\usepackage[english]{babel} 
\usepackage{blindtext} 
\usepackage{ifthen} 
\usepackage{orcidlink} 
\usepackage{physics}
\usepackage[normalem]{ulem}
\graphicspath{{figures/}} 

\newboolean{articletitles}
\setboolean{articletitles}{true} 

\input{belle2-symbols}

\usepackage{verbatim}
\usepackage{dcolumn}
\usepackage{amsmath}
\usepackage{epsfig}
\usepackage{subfigure}
\usepackage{xcolor}

\newcommand{\ktt}{\ensuremath{B^{+} \to K^{+} \tau^{+} \tau^{-}}\xspace}
\newcommand{\eecl}{\ensuremath{E_{\text{extra}}}\xspace}

\def\epem{e^+e^-}

\newcommand{\plcms}{\ensuremath{p^{*}_{\ell^{+}}}\xspace}
\def\Btag{\ensuremath{B_{\rm tag}}\xspace}

\def\BpBm    {\ensuremath{\Bu {\kern -0.16em \Bub}}\xspace}

\newboolean{wordcount}
\setboolean{wordcount}{false} 

\ifthenelse{\boolean{wordcount}}%
{\usepackage{bibentry} 
 \usepackage{comment} 
 \excludecomment{align*} 
 \excludecomment{align} 
 \excludecomment{equation*} 
 \excludecomment{equation} 
 \excludecomment{eqnarray*} 
 \excludecomment{eqnarray} 
 \excludecomment{acknowledgments} 
 \def\maketitle{} 
}{}

\begin{document}

\begin{flushright}
Belle II preprint 2026-003 \\
KEK preprint 2025-43 \\     
\end{flushright}

\title{
 Search for the decay $B^+ \rightarrow K^+\tau^+\tau^-$ using data from the Belle and Belle~II experiments}

\ifthenelse{\boolean{wordcount}}{}{
  \author{M.~Abumusabh\,\orcidlink{0009-0004-1031-5425}} 
  \author{I.~Adachi\,\orcidlink{0000-0003-2287-0173}} 
  \author{K.~Adamczyk\,\orcidlink{0000-0001-6208-0876}} 
  \author{A.~Aggarwal\,\orcidlink{0000-0002-5623-3896}} 
  \author{L.~Aggarwal\,\orcidlink{0000-0002-0909-7537}} 
  \author{H.~Ahmed\,\orcidlink{0000-0003-3976-7498}} 
  \author{Y.~Ahn\,\orcidlink{0000-0001-6820-0576}} 
  \author{H.~Aihara\,\orcidlink{0000-0002-1907-5964}} 
  \author{N.~Akopov\,\orcidlink{0000-0002-4425-2096}} 
  \author{S.~Alghamdi\,\orcidlink{0000-0001-7609-112X}} 
% ...
\collaboration{The Belle and Belle II Collaborations}

}

\begin{abstract}
We report a search for the rare decay \Bp\to$K^{+}$\tautau using $1.2 \times 10^9$ \FourS mesons produced near threshold in electron-positron collisions  and collected by the Belle and Belle~II experiments.  We fully reconstruct the hadronic decay of one $B$ meson produced in the $\ensuremath{\Upsilon{(4S)}}\rightarrow B^{+} B^{-}$ decay, and search for $B^{\pm}\rightarrow K^{\pm} \tau^{+}\tau^{-}$ candidates among the remaining collision products, reconstructing a charged kaon and leptonic decays of the $\tau$ leptons.  We optimize the selection for best sensitivity and look for an excess over background at low values of the residual energy detected in the calorimeter after full event reconstruction. We observe no significant excess and set the limit \mbox{$\mathcal{B}(B^{+}\rightarrow K^{+}\tau^{+}\tau^{-})< 0.56\times 10^{-3}$} at the 90\% confidence level, improving on the only previous result by a factor of four.
\end{abstract}

\maketitle


Identifying the particles and interactions that extend the standard model (SM) is currently a principal objective of particle physics. 
Weak-interaction processes that change quark flavor without changing the electric charge, known as flavor-changing neutral currents, are promising probes for non-SM physics. In the SM, the amplitudes for these processes are suppressed, because they cannot occur via single $W^\pm$ emission and require a loop-mediated $W^\pm$ exchange. However, virtual non-SM particles of high mass can replace the exchanged SM particles in those loop amplitudes. This could alter the process rates, indicating non-SM physics.  Among flavor-changing neutral current processes, the $B^+ \to K^+\tau^+\tau^-$ decay is particularly interesting. The SM branching fraction, subtracting $B^+ \to K^+[c\bar{c}\to]\tau^+\tau^-$ contributions, is expected to be (1.0--$2.0) \times 10^{-7}$~\cite{Bouchard:2013mia, hewitt, Parrott:2022zte}. However, non-SM contributions proposed~to accommodate the observed excesses in $B^{+,0} \to \overline{D}{}^{0,-(*)} \tau^+  \nu_\tau$~\cite{HeavyFlavorAveragingGroupHFLAV:2024ctg} and $B^{+} \to K^{+} \nu \bar{\nu}$~\cite{Belle-II:2023esi} decay rates could enhance it by a factor of up to $10^3$~\cite{Alonso:2015sja, Barbieri:2015yvd, Capdevila:2017iqn, Kumar:2018kmr,Allwicher:2023xba}, bringing it near the current experimental sensitivity. 

The only search for $B^{+} \to K^{+} \tau^{+} \tau ^{-}$ decays to date was reported by the \babar{} experiment, which set a 90\%-confidence-level (CL) upper limit on the branching fraction at $2.25 \times 10^{-3}$~\cite{BaBar:2016wgb}. Null searches for related processes have also been reported by the Belle, Belle II, and LHCb experiments~\cite{Belle:2021ecr,LHCb:2017myy, Belleii:moneta, LHCb:2026kstt}. All the resulting bounds exceed the SM predictions by three to four orders of magnitude, leaving opportunities for further experimental exploration. 

This Letter reports a search for the rare decay \Bp\to$K^{+}$\tautau~\cite{Vobbilisetti:2023fnz, debjit:thesis}. Charge-conjugate processes are implied throughout. The search uses $772 \times 10^6$ \FourS meson decays produced near threshold in electron-positron collisions, and reconstructed from the full  711~fb$^{-1}$ data set collected by the Belle experiment, and $387\times 10^6$ \FourS decays from the 365~fb$^{-1}$ data set collected by the Belle~II experiment between 2019 and 2023. The collisions are produced by the KEKB and SuperKEKB asymmetric-energy colliders at KEK near the center-of-mass (c.m.) energy $10.58~\mathrm{GeV}$, which corresponds to the peak of the $\Upsilon(4 S)$ resonance. 

The principal challenge in this analysis arises from the presence of multiple final-state neutrinos from the $\tau$-lepton decays. The neutrinos are not reconstructed, precluding the availability of distinctive signal kinematic distributions needed to suppress $10^8$-times larger backgrounds. The analysis therefore exploits the advantages of near-threshold $B\overline{B}$ pair production to identify signal.  We infer the signal four-momentum and suppress background by fully reconstructing a hadronic decay of the partner $B^-$ meson, pair-produced with the signal $B^+$ meson in the $e^{+} e^{-} \rightarrow \Upsilon(4 S) \rightarrow B^+ B^-$ process.  We reconstruct signal candidates in events containing a kaon candidate with charge opposite to the partner-$B$ charge and two opposite-charge lepton candidates. Using only decays $\tau^+\to \ell^+ \nu_{\ell} \bar{\nu}_{\tau}$  of the signal $\tau$ leptons, where $\ell$ denotes a muon or an electron, further reduces background and simplifies the analysis without degrading sensitivity.  A restriction to events with opposite-charge kaon-lepton mass larger than the charmed-meson mass suppresses large backgrounds dominated by $b \to c (\to s \ell^{(\prime)}\nu)  \ell \overline{\nu}$ quark-level processes.  Near-threshold $B^+B^-$ production implies also that no additional particles, and therefore no additional energy in the calorimeter, are expected after $B^+B^-$ reconstruction. Hence, we determine the signal yield by counting the excess events, over the expected background yield, at low values of the calorimeter energy detected after subtraction of the observed energies of both $B$ meson candidates. The analysis is executed using simulated and control-data samples before examining the data in the signal-search region. The search is conducted independently in the Belle and Belle~II data, and the results are combined.


The Belle detector~\cite{Abashian2002117} collected data from collisions of 8-GeV electrons with 3.5-GeV positrons produced by the KEKB collider~\cite{Kurokawa:2001nw}. An upgraded detector, Belle~II~\cite{Abe:2010gxa}, studies collisions of 7-GeV electrons on 4-GeV positrons produced by the SuperKEKB collider~\cite{Akai:2018mbz}. Each detector is a nearly $4\pi$ hermetic solenoidal magnetic spectrometer comprising inner silicon-vertex subdetectors and an outer drift chamber, surrounded by Cherenkov-based charged-particle identification subdetectors, a crystal electromagnetic calorimeter, and outer subdetectors for penetration-based detection of muons and $K^0_L$ mesons.

Detailed Monte Carlo simulation is used to design the analysis, decide selection criteria and determine their efficiencies, study backgrounds, and assess the analysis sensitivity. Simplified simulations are used to study estimator properties. The software packages {\sc EvtGen}~\cite{Lange:2001uf}, {\sc PYTHIA6}~\cite{Sjostrand:2006za}~({\sc PYTHIA8})~\cite{Sjostrand:2014zea}, {\sc KKMC}~\cite{Jadach:1999vf}, and {\sc TAUOLA}~\cite{Jadach:1990mz} simulate particle production and decay in Belle (Belle II); {\sc PHOTOS}~\cite{Barberio:1990ms} simulates photon radiation from final-state charged particles, and {\sc GEANT3}~\cite{Brun:1987ma}~({\sc GEANT4}~\cite{Agostinelli:2002hh}) simulates interactions with the Belle (Belle~II) detector and its response. We generate $5\times 10^7$ $e^+e^- \to \FourS \to B^+(\to K^+\tau^+\tau^-)B^-$ signal events for each experiment. The decay of the partner bottom meson is unbiased. The signal decay follows the dynamical model of Ref.~\cite{Ali:2000btosll}. 
Simulated background samples are four times larger than the collision-data sample. They consist of \BpBm events, $B^0$\Bzb events, and continuum $e^{+} e^{-} \rightarrow q \bar{q}$ events, where $q$ stands for a $u$, $d$, $s$, or $c$ quark. None of the simulated $e^+e^- \to \tau^+\tau^-$ events passes the selection. We use the Belle~II software~\cite{Kuhr:2018lps, basf2-zenodo} to reconstruct simulated and collision events in both Belle~\cite{b2bii} and Belle~II data. 

A trigger selects events online based on total energy and neutral-particle multiplicity to preferentially retain collisions that yield hadrons, with full efficiency for signal events. 

In the offline analysis,  we fully reconstruct the hadronic decay of the candidate partner $B$ meson using the full-event interpretation algorithm (FEI)~\cite{Keck:2018lcd}. The FEI selects final-state particles originating from the interaction space-point through standard quality criteria and combines them bottom-up according to known hadronic decay chains to achieve full $B$ meson reconstruction. Backgrounds are suppressed by means of two kinematic observables that exploit four-momentum conservation in near-threshold $B\overline{B}$ pair production,
$M_{\mathrm{bc}} = \sqrt{(E^*_{\rm beam})^2/c^4 - |\vec{p}^{\, *}_{B}|^2/c^2}$ and 
 $\Delta E  = E^*_{B} -  E^*_{\rm beam}$,
where $E^*_{\rm beam}$ is the beam energy and $(E^*_{B}/c,\vec{p}^{\,*}_{B})$ is the four-momentum of the $B$ candidate, both calculated in the c.m.\ frame. The $M_{\mathrm{bc}}$ and $\Delta E$ distributions for correctly reconstructed $B$ decays peak at the $B$ mass and close to zero, respectively. 
Hence, partner $B$ mesons are required to have $M_{\mathrm{bc}}  > $ 5.27~\gevcc and $-0.15  < \Delta E < 0.10~\gev$. Furthermore, we require that only three charged particles are reconstructed in addition to the partner $B$ and impose a loose restriction on the FEI-reconstruction quality. 
We use topological observables that exploit the differences between the jet-like distribution of continuum products and the isotropic distribution of $B\overline{B}$ decay products to reduce the large
continuum background~\cite{BaBar:2014omp}: the spatial distributions of particle momenta must be isotropic~\cite{Bjorken:1969wi} and the thrust direction of the partner $B$ not opposite to the thrust direction of all other particles in the event.
Signal-candidate selections reduce continuum background by 68\%~(65\%) in Belle (Belle~II) data while reducing signal efficiency by 15\% (13\%). If multiple partner $B$ candidates are reconstructed in an event, the candidate with highest FEI quality is chosen. Partner $B$ reconstruction retains 0.8\%~(0.5\%) of Belle (Belle~II) simulated signal events and the partner $B$ decay is correctly reconstructed in 28.7\%~(43.9\%) of the simulated signal events.


We use the three charged particles not associated with the partner $B$ to reconstruct signal $B$ candidates. We select only charged-particle trajectories (tracks) loosely originating from the interaction space-point to suppress spurious tracks from beam-induced background.  
Furthermore, we require tracks to be reconstructed within the polar acceptance of the drift chambers and to either be associated with a sufficient number of chamber measurement points, in Belle~II data, or to correspond to particles with momentum $p_T>200~\mathrm{MeV} / c$ transverse to the beam direction, in Belle data. This ensures reliable measurements of specific-ionization energy losses for charged-particle identification (PID). 
Kaons, pions, muons, and electrons are identified using PID information from several subdetectors combined into likelihood-ratio-like observables.  
Muon identification is performed using only information from the muon subdetector for Belle data, and using all subdetectors for Belle~II data. This implies that Belle muon candidates have transverse momenta larger than $0.6~\mathrm{GeV} / c$, as needed to reach the outer detectors, resulting in smaller efficiency and misidentification rate relative to Belle II data. The PID criteria yield average efficiencies of 92.1\%  (90.0\%) for kaons, 74.7\% (83.0\%) for electrons, and 43.6\% (65.0\%) for muons, and fractions of 4.6\% (4.5\%), 0.2\% (2.0\%), and 0.8\% (2.7\%) of misidentified candidates in Belle (Belle~II) data, as determined in control samples as functions of momentum and polar angle~\cite{Belleii:hadronid}.

The reconstruction of calorimeter energy deposits (clusters) generated by neutral particles is critical, as signal extraction relies on those clusters. To suppress beam-induced background photons, we select clusters with energies exceeding, respectively, 100.0 (100.0)~MeV, 50.0 (55.0)~MeV, and 150.0~(150.0)~MeV in the forward, central, and backward (with respect to the electron-beam direction) regions of the calorimeter for Belle~(Belle~II) data. In Belle~II data, we restrict the cluster polar angle to the drift-chamber acceptance to suppress electrons misreconstructed as photons. Cluster selections in Belle~II are more stringent than in Belle to suppress higher beam-induced backgrounds. We also require clusters to be reconstructed at least 30 cm from the closest extrapolated
track to suppress background due to secondary particles generated in nuclear interactions of charged hadrons with the calorimeter. These criteria minimize data-to-simulation differences in cluster-related quantities as observed in  several control samples~\cite{debjit:thesis}.


A signal $B$-meson candidate is formed by combining a kaon candidate of charge opposite to that of the partner $B$ meson with two opposite-charge leptons in $e^+e^-$, $e^\pm\mu^\mp$, or $\mu^+\mu^-$ combinations. We analyze these three dilepton combinations together, as simulation shows no significant sensitivity gain from separate analyses. We suppress $B^+ \to J/\psi(\to \ell^+\ell^-)K^+$ background by restricting the dilepton mass to be less than $3.0$~GeV/$c^2$, which is the region relevant for our signal. We also  
suppress $\gamma\to\epem$ conversion backgrounds by requiring the electron-positron mass to exceed 100~\mevcc. 

We then combine signal and partner $B$ candidates to form $\Upsilon(4S)$ candidates. For each $\Upsilon(4S)$ candidate, we construct the squared momentum transfer (or beam-constrained squared $\tau$-pair mass) $q^2 = (p_{e^+e^-} - p_{\Btag} -p_K)^2$, where $p_{{e^+e^-}}$, $p_{{ \Btag}}$, and  $p_{{K}}$ are the sum of the four-momenta of both beams, the four-momentum of the partner $B$, and that of the kaon, respectively. We impose $q^2 > 14.18$~GeV$^2$/$c^4$, corresponding to the signal kinematic acceptance used in SM predictions~\cite{Bouchard:2013mia}. This incidentally suppresses  contributions from $B^+ \to \psi(2S) (\to \tautau) K^+$ decays, whose rate is ten-times larger than the expected SM-signal. We further reject background by vetoing events that have \piz candidates  unassociated with the $\Upsilon(4S)$ reconstruction. These candidates are identified as pairs of neutral clusters detected in the central calorimeter region with diphoton masses between 131 and 138~\mevcc. These selections remove 99\%~(99\%) of the residual background after the partner $B$ selection with 3.1\%~(5.8\%) signal efficiency in Belle~(Belle~II) data.

\begin{figure}
\centering
\begin{tabular}{c}
\includegraphics[width=0.8\linewidth]{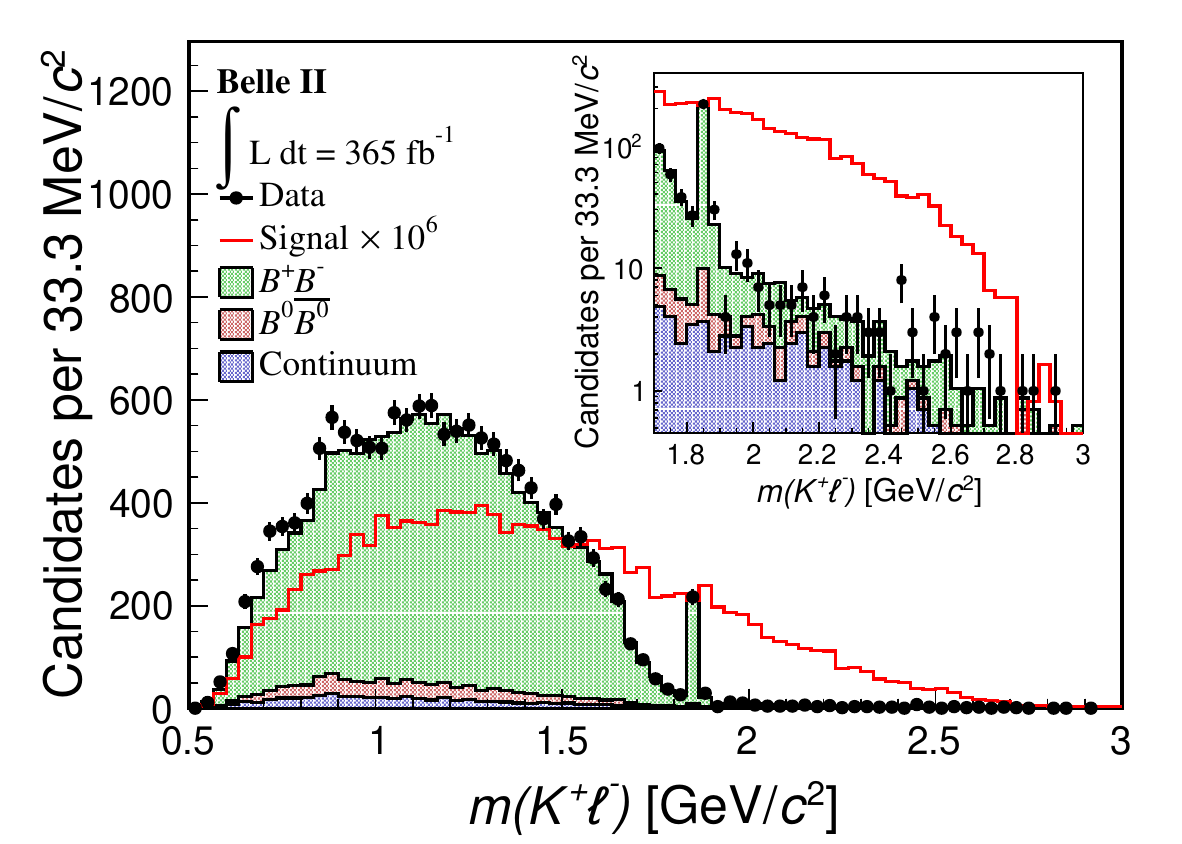} \\
\end{tabular}
\caption{Distribution, with simulation overlaid, of opposite-charge kaon-lepton mass in Belle~II events. Logarithmic vertical scale in the inset assists visualization of the high kaon-lepton mass analysis region.}
 \label{fig:mkl}
\end{figure}

At this stage, remaining backgrounds are primarily $B^+B^-$ events in which a properly reconstructed partner $B^-$ is accompanied by a $B^+ \to \overline{D}{}^{0}(\to K^+ \ell^{(\prime)-} \overline{\nu}_{\ell^{(\prime)}}) \ell^{+} \nu_{\ell}$ decay or the analogous decay involving a $\overline{D}{}^{*0}$ instead of a $\overline{D}{}^{0}$ meson, which are both signal-like. The opposite-charge kaon-lepton mass $m(K^+\ell^-)$ exhibits a major transition in background size and composition at the $D^0$ meson mass, as shown in Fig.~\ref{fig:mkl}. The narrow peak is from $D^0 \to K^-\pi^+$ decays with the pion misidentified as a muon. The large component of $B^+$ mesons decaying into $D$ mesons that in turn decay into $\pi^{0}$'s unassociated with the candidate populates masses lower than the charm mass. This results in a significantly higher signal-to-background ratio at higher $m(K^+\ell^-)$ values. Restricting the search to events with $m(K^+\ell^-) > 1.9$~GeV/$c^2$ reduces the remaining background by 98.7\%~(99.2\%) while keeping 20\%~(16\%) of signal in Belle (Belle~II) data, thus constituting an important analysis choice. 

The principal remaining backgrounds are $B \to D^{(\ast)} \ell^{+} \nu$ decays followed by semileptonic $D$ decays and continuum events. 
These are further suppressed through an optimized selection on the following three additional discriminating quantities. 
The squared missing mass $M_{\rm miss}^2= (\textit{p}_{e^+e^-} - \textit{p}_{\rm{tot}})^{2}/c^2$, where $p_{\text{tot}}$ is the total four-momentum of all particles reconstructed in the event, is sensitive to the higher number of neutrinos in signal events compared to background events~\cite{SuppMat}. The c.m.\ momentum \plcms of the signal lepton with same charge as the signal kaon also has a distinctive spectrum for signal. 
The third variable is the search-region width in the signal-extraction observable $E_{\text {extra}}$, which is the summed energy of all neutral clusters not used in \FourS reconstruction.  
The selection requirements for these are optimized to provide the best average expected (statistical only) upper limit in background-only simulation, yielding $M^2_{\rm miss} > 3.0~(1.6)$~GeV$^2/c^4$, $\plcms > 0.4~(0.5)$ GeV/$c$, and $E_{\text {extra }}<100~(250)$ \mev for Belle~(Belle~II). No event contains more than one reconstructed signal $B$ candidate after the selection.

\begin{figure}
\centering
\begin{tabular}{c}
\includegraphics[width=0.8\linewidth]{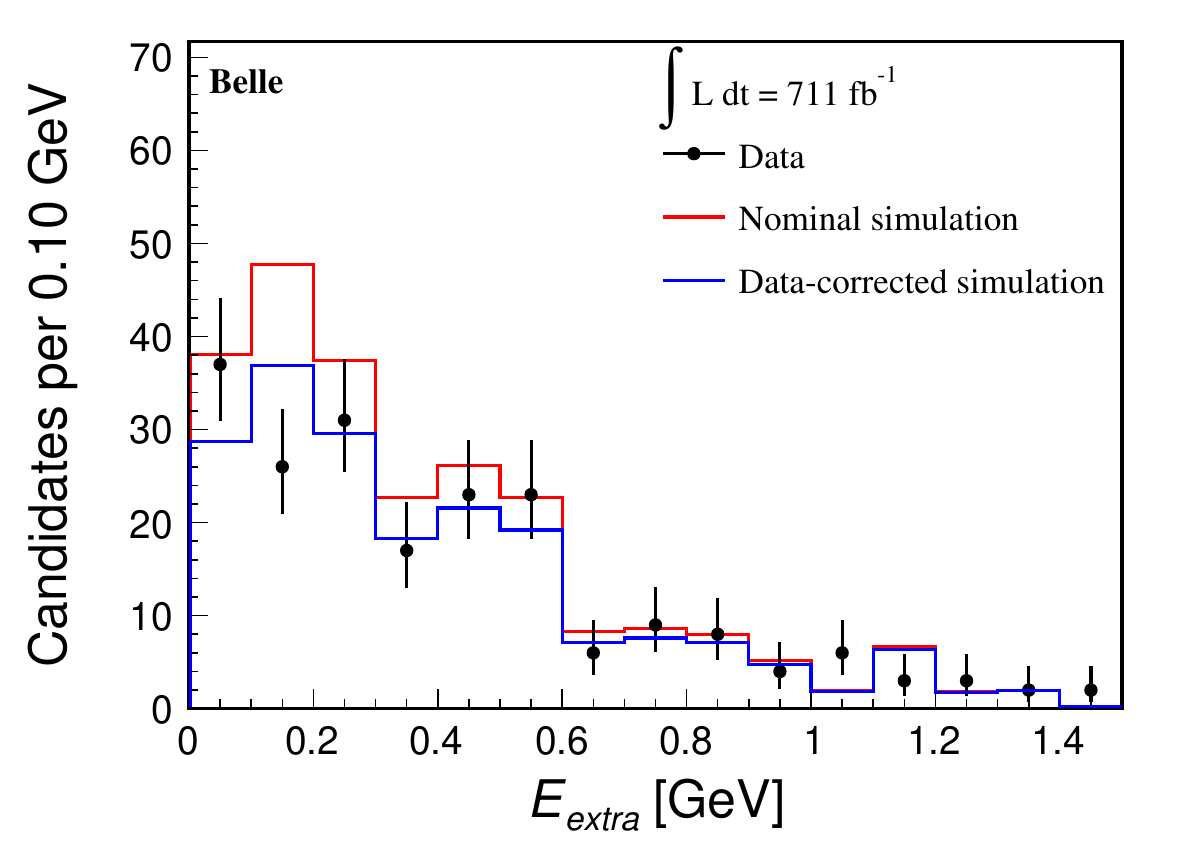} \\
\includegraphics[width=0.8\linewidth]{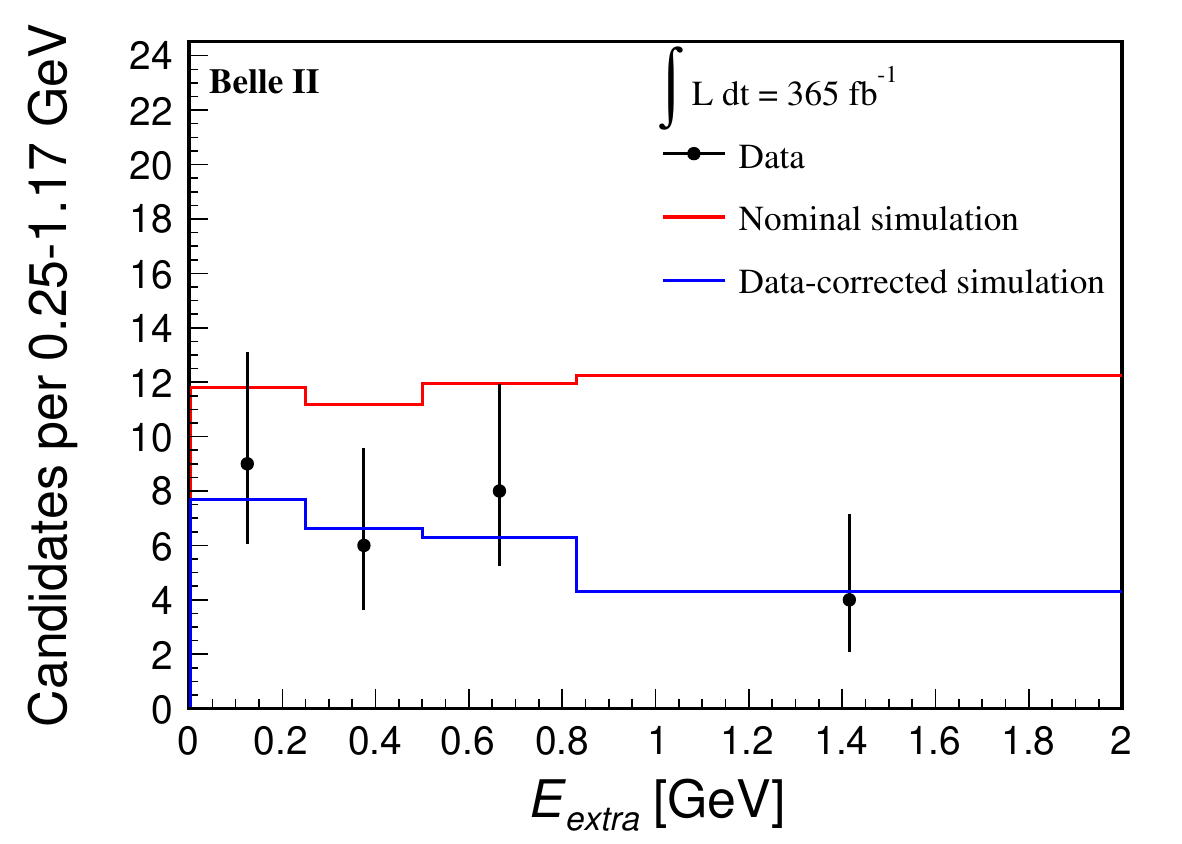}\\
\end{tabular}
\caption{Distributions of \eecl for (circles) data and (histograms) simulation (red) before and (blue) after the linear-scaling fit for (top) Belle and (bottom) Belle~II events populating the partner-$B$ $\rm M_{bc}$ sideband.}
 \label{fig:optifitprojection}
\end{figure}

We determine the signal yield by counting the excess events over the background expectation, $N_{\rm sig} = N_{\rm obs} - N_{\rm bkg}$, in the signal-search $E_{\text{extra}}$  region. The background expectation is obtained by extrapolating, in simulation, the background yields observed in three independent signal-depleted sidebands. These are events with \( q^2 < 14.18 \)~GeV$^2$/$c^4$, events in the \( 5.20 < M_{\text{bc}} < 5.27 \)~GeV/\(c^2\) sideband of the partner $B$, and events with \( E_{\text{extra}} > 100~(250) \)~MeV for Belle~(Belle~II) data. For each of these sidebands, the selection on the other two observables is the same as for the signal-search region. We validate the simulated $E_{\text{extra}}$ distribution by comparing it with experimental data in the three sidebands.
Good agreement is observed, but within large statistical uncertainties due to small sample sizes. Hence, we further ensure reliability of the background extrapolation into the signal-search region by correcting the simulated $E_{\text{extra}}$ shape and normalization in those sidebands to match the data.  A likelihood-ratio test~\cite{kendall:book} favors
a scaling between data and simulation based on a linear function of $E_{\text{extra}}$. The scaling parameters are determined with a likelihood fit to the binned ratios of data and simulation distributions, simultaneous across the three sidebands (Fig.~\ref{fig:optifitprojection} Ref.~\cite{SuppMat}). The distribution is Poisson in each bin, for both data and simulated-event counts. We then extrapolate the corrected simulation into the signal-search $E_{\text{extra}}$ region to determine the expected background yields, $14.1 \pm 1.6$ events for Belle and $3.5 \pm 0.7$ events for Belle~II, where the uncertainties are only statistical.

%

\begin{figure}
    \centering
    \begin{tabular}{c}
     \includegraphics[width=0.8\linewidth]{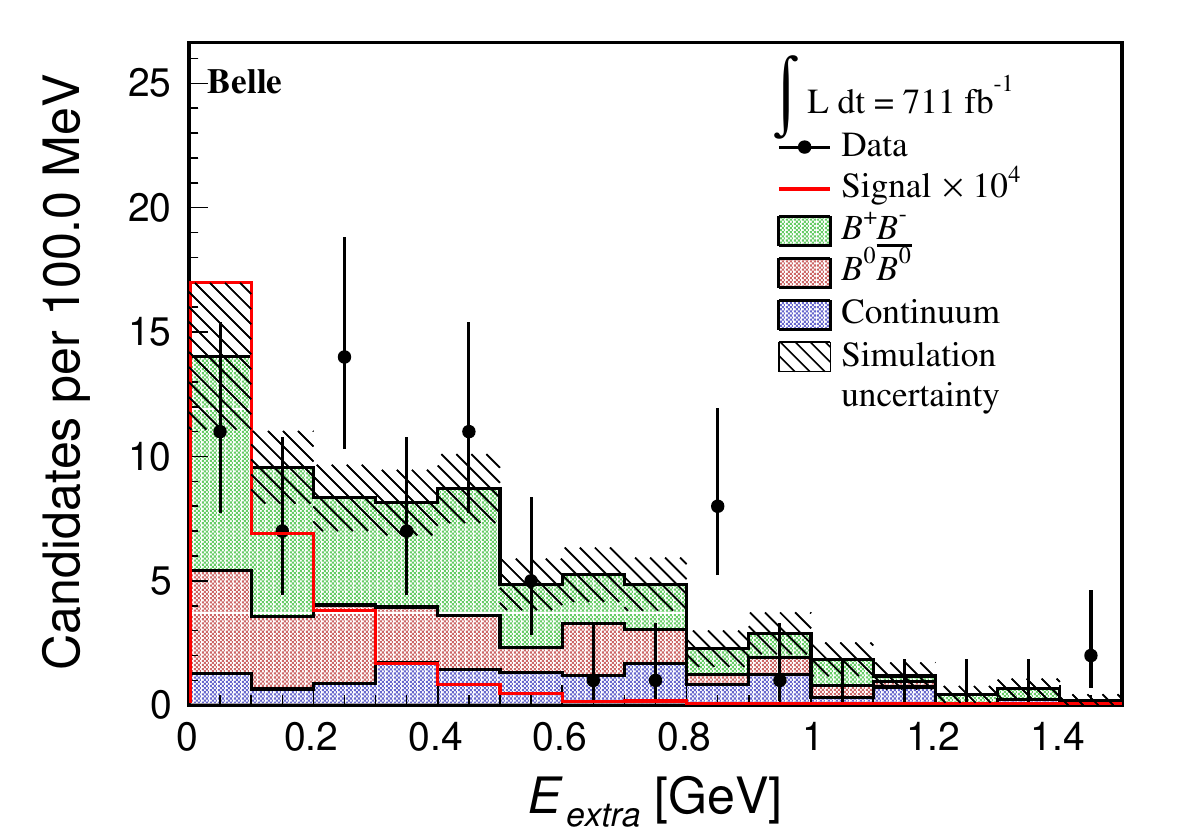} \\
     \includegraphics[width=0.8\linewidth]{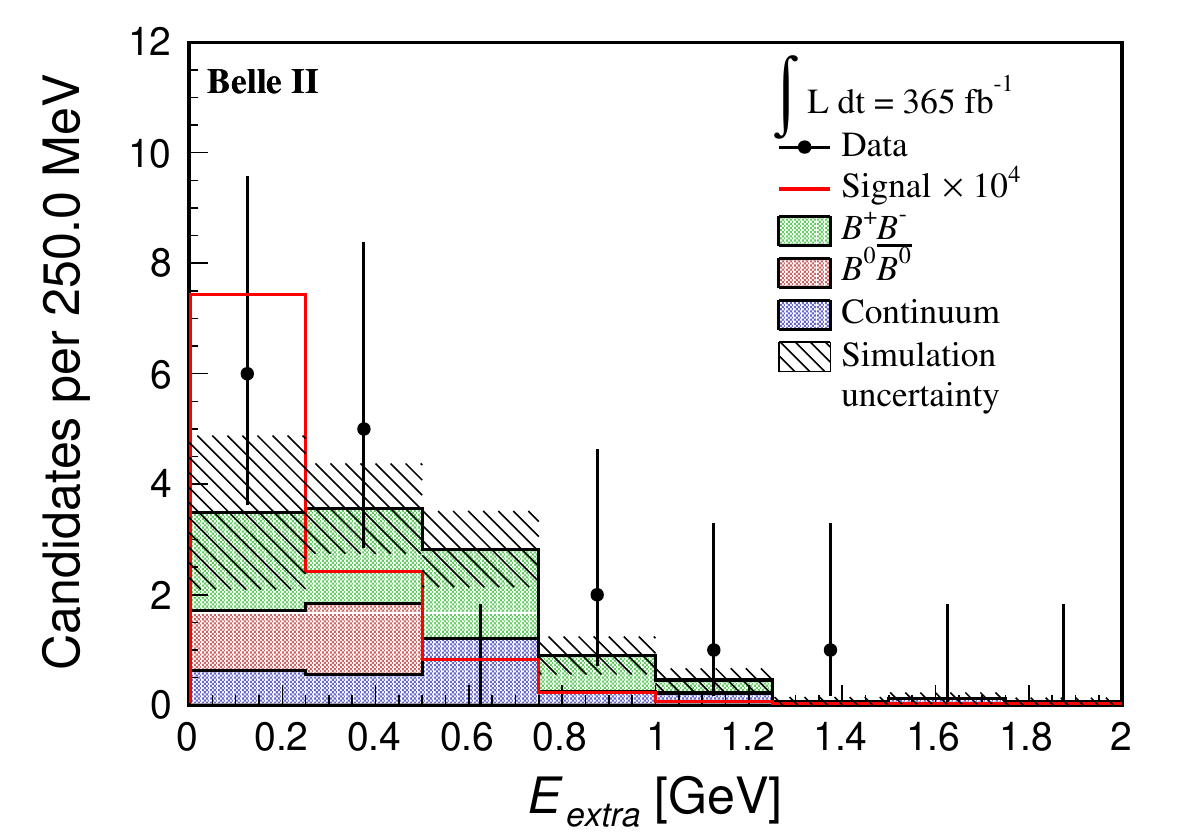} \\
    \end{tabular}
    \caption{Distributions of $E_{\text {extra }}$ after final selections in (top) Belle and (bottom) Belle~II data. Expectations based on corrected simulation for backgrounds and for $10^4\times$-enhanced SM signal~\cite{Bouchard:2013mia} are overlaid.}
    \label{fig:eecl_data}
\end{figure}

Figure~\ref{fig:eecl_data} shows the distributions of $E_{\text {extra}}$ in the data after final selections, with background expectations from simulation overlaid. The first bin of each distribution is the signal-search region. The event yields observed in the Belle and Belle~II signal-search regions, 11 and 6, respectively, show no significant excesses with respect to the expectations. 

The signal branching fraction is determined as
\begin{equation}
\mathcal{B}(\ktt) =  \frac{N_{\rm sig}}{2 \epsilon f^{+-}  N_{\Upsilon(4S)}},
\label{eq:bf}
\end{equation}
where 
$\epsilon$ is the efficiency for reconstructing signal, which includes the relevant product of ${\cal B}(\tau \to \ell \nu_\tau)$ factors that are $(17.81\pm 0.04)\%$ for electrons and $(17.39\pm 0.04)\%$ for muons, and $f^{+-}=0.511 ^{+0.007} _{-0.011}$ is the $B^+B^-$ fraction~\cite{HeavyFlavorAveragingGroupHFLAV:2024ctg} of the $N_{\FourS}= [(772\pm 11)+ (387\pm 6)] \times 10^{6}$ \FourS decays recorded in the combined data set.
Efficiencies for signal are determined from simulation and corrected using control data. We use $B^+\to \overline{D}{}^{0}\pi^+$ decays to match to data the efficiency for the partner-$B$ FEI reconstruction in simulation.  We reconstruct partner $B$ candidates using FEI and select 1.5--3.0-GeV/$c$-momentum charged pions in the rest of the event. We fit the yields of charmed-meson signals appearing in the spectrum of recoil mass, $\sqrt{(E^{*}_{\mathrm{beam}} - E^{*}_{\pi})^{2}/c^4 - (\vec{p}^{\, *}_{B} +  \vec{p}^{\, *}_{\pi} )^{2}/c^2}$, where $E^{*}_\pi$ and $\vec{p}^{\, *}_\pi$ are the energy and momentum of the pion and $\vec{p}^{\, *}_B$ is the momentum of the partner $B$ in the c.\m.\ frame~\cite{Vobbilisetti:2023fnz}. We then compare the resulting yields with those observed in simulation and use their ratio $r_{\rm FEI} = 0.76 \pm 0.03~(0.74 \pm 0.05)$ to correct the Belle (Belle~II) FEI efficiency observed in simulation. We use analysis candidates populating the 1.8--1.9 GeV/$c^2$ range in the opposite-charge kaon-lepton mass to determine the $\pi^0$ veto efficiency. This sample is dominated by $B^+ \to \overline{D}{}^0(\to K^+\pi^-)\ell^+ \nu$ decays in which the pion is misreconstructed as a lepton, yielding a narrow peak. A simultaneous fit to the events that pass and those that fail the $\pi^0$ veto determines the efficiency ratio between data and simulation. This ratio, $1.03 \pm 0.02$~($1.03 \pm 0.03$) for Belle~(Belle~II) corrects the $\pi^0$ veto efficiency in simulation, assuming similar $\pi^0$ multiplicities in data and simulation.

The primary source of systematic uncertainty is the additive uncertainty on the estimate of the background yield in the signal region, which is 17.4\%~(33.6\%), relative to the central value of the background estimate, for Belle (Belle~II).  This combines the 9\% (18\%) statistical uncertainty on the extrapolation, due to the finite sizes of the sideband samples, and the systematic uncertainty associated with the shape assumed for the background $E_{\text {extra}}$ scaling. The latter is assessed by repeating the extrapolation with a uniform or a quadratic scaling with respect to $E_{\text {extra}}$ and taking the largest difference in signal yield with respect to the default linear scaling. The remaining systematic uncertainties are all multiplicative. The second-largest contribution is the uncertainty on the partner-$B$ reconstruction efficiency. This combines the uncertainty on the data-to-simulation efficiency ratio and the efficiency change observed when using an alternative approach to calibrate the FEI reconstruction efficiency, based on inclusive $B^+ \to X_c \ell^+ \nu$ decays reconstructed using only the lepton. The systematic uncertainty on the signal branching fraction is 10.0\% (13.3)\% for Belle~(Belle~II). The uncertainties on PID-related efficiencies in Belle data are determined using $J / \psi \rightarrow \ell^{+} \ell^{-}$and $D^{*+} \rightarrow D^0\left(\rightarrow K^{-} \pi^{+}\right) \pi^{+}$ decays to be $2.4\%, 2.5\%$, and $1.6\%$ for muons, electrons, and kaons, respectively. The corresponding Belle~II uncertainties are 0.6\%, 0.3\%, and 0.9\%, obtained using the same processes, as well as $e^+e \to \ell^+ \ell^- e^+e^-$ and $e^+e^- \to \ell^+ \ell^-\gamma$. The uncertainty for the $\pi^0$ veto efficiency is $2.2\%~(2.5)\%$ in Belle (Belle~II) data, as determined from $B^+ \to \overline{D}{}^{0}(\to K^+\pi^-)\ell^+\nu$ decays reconstructed as a narrow peak in the opposite-charge kaon-lepton mass. An uncertainty of $0.35 \%$~\cite{BaBar:2014omp:sec15} ($0.27 \%$) per charged particle is assigned to account for possible data-simulation differences in the Belle~(Belle~II) tracking efficiency. This uncertainty is fully correlated between the three signal tracks, leading to a $1.05 \%$ ($0.81 \%$) uncertainty on the signal branching fraction. We use a simulated $B^+ \rightarrow K^{+} \tau^{+} \tau^{-}$ sample generated with an alternative model for the hadronic form-factors~\cite{Parrott:2022zte}, to determine a 3.8\%~(4.0\%) uncertainty due to the SM signal modeling. The systematic uncertainties due to the limited sizes of simulated samples used to determine efficiencies are 3.1\%~(3.3\%) for Belle~(Belle II). The uncertainties arising from the values of $N_{\FourS}$ and $f^{+-}$ are $1.4\%~(1.6\%)$ and $ ^{+1.4}_{-2.1}\%$, respectively, for Belle~(Belle~II).  


We determine exclusion limits using the CLs method based on a profile-likelihood-ratio test~\cite{Junk:1999, A_L_Read_2002}. Systematic uncertainties are incorporated with Gaussian constraints. Coverage tests show no significant undercoverage. Table~\ref{table1} shows the results of the search, which is conducted independently in the Belle and Belle~II samples. The resulting branching fractions are averaged yielding $\mathcal{B}(B^+ \to K^+\tau^+\tau^-)= (0.04^{+0.27 +0.17}_{-0.23 -0.18}) \times 10^{-3}$. The corresponding combined exclusion limit is \mbox{$\BR(\Bp\to K^{+}\tautau) < 0.56 \times 10^{-3}$} at the 90\% confidence level.

\begin{table}
\begin{center}
\begin{tabular}{lccc}
\hline
\hline
 & Belle & Belle~II \\ \hline 
$N_{\rm bkg}$   & $14.1 \pm 1.6 \pm 1.9$ & $3.5 \pm 0.7 \pm 0.9$  \\
$N_{\rm obs}$     & 11 & 6 \\ 
$\epsilon \times 10^{5}$ & $1.45 \pm 0.05 \pm 0.18$ & $1.26 \pm 0.04 \pm 0.18$\\
${\cal B} (B^+ \to K^+\tau^+\tau^-)\times 10^{4}$ & $-2.7^{+3.2}_{-2.6}\pm {2.2}$ & $5.1^{+5.6}_{-4.3}\pm {2.5}$\\
Obs.~(exp.) limit ($10^{-3}$)& 0.5$~(0.7)$ & 1.4$~(0.9)$ \\

\hline
\hline
\end{tabular}
\caption{Expected background yields, observed event yields, signal efficiencies, observed branching fractions, and observed~(expected) 90\% CL exclusion limits.}
\label{table1}
\end{center}
\end{table}

In summary, we search for the rare decay \mbox{ \Bp\to$K^{+}$\tautau} using $772 \times 10^6$ \FourS meson decays from the full data set of electron-positron collisions at the \FourS resonance collected by the Belle experiment and $387\times 10^6$  \FourS decays collected by the Belle~II experiment between 2019 and 2023. The analysis achieves world-leading results by means of an optimized selection applied to a large data set. We observe no significant signal and set a 90\% CL upper limit at \mbox{$\BR(\Bp\to K^{+}\tautau) < 0.56 \times 10^{-3}$}. This limit is four times more stringent than the only previous bound and contributes constraints to the parameters of several SM extensions~\cite{Alonso:2015sja, Barbieri:2015yvd, Capdevila:2017iqn, Kumar:2018kmr,Allwicher:2023xba}.

\begin{acknowledgments}
This work, based on data collected using the Belle II detector, which was built and commissioned prior to March 2019,
and data collected using the Belle detector, which was operated until June 2010,
was supported by
Higher Education and Science Committee of the Republic of Armenia Grant No.~23LCG-1C011;
Australian Research Council and Research Grants
No.~DP200101792, 
No.~DP210101900, 
No.~DP210102831, 
No.~DE220100462, 
No.~LE210100098, 
and
No.~LE230100085; 
Austrian Federal Ministry of Education, Science and Research,
Austrian Science Fund (FWF) Grants
DOI:~10.55776/P34529,
DOI:~10.55776/J4731,
DOI:~10.55776/J4625,
DOI:~10.55776/M3153,
and
DOI:~10.55776/PAT1836324,
and
Horizon 2020 ERC Starting Grant No.~947006 ``InterLeptons'';
Natural Sciences and Engineering Research Council of Canada, Digital Research Alliance of Canada, and Canada Foundation for Innovation;
National Key R\&D Program of China under Contract No.~2024YFA1610503,
and
No.~2024YFA1610504
National Natural Science Foundation of China and Research Grants
No.~11575017,
No.~11761141009,
No.~11705209,
No.~11975076,
No.~12135005,
No.~12150004,
No.~12161141008,
No.~12405099,
No.~12475093,
and
No.~12175041,
and Shandong Provincial Natural Science Foundation Project~ZR2022JQ02;
the Czech Science Foundation Grant No. 22-18469S,  Regional funds of EU/MEYS: OPJAK
FORTE CZ.02.01.01/00/22\_008/0004632 
and
Charles University Grant Agency project No. 246122;
European Research Council, Seventh Framework PIEF-GA-2013-622527,
Horizon 2020 ERC-Advanced Grants No.~267104 and No.~884719,
Horizon 2020 ERC-Consolidator Grant No.~819127,
Horizon 2020 Marie Sklodowska-Curie Grant Agreement No.~700525 ``NIOBE''
and
No.~101026516,
and
Horizon Europe Marie Sklodowska-Curie Staff Exchange project JENNIFER3 Grant Agreement No.~101183137 (European grants);
L’Institut National de Physique Nucl\'eaire et de Physique des
Particules (IN2P3) du CNRS under Project Identification No.
CNRS-IN2P3-14-PP-033
and L’Agence Nationale de la Recherche (ANR) under Grant No. ANR-23-CE31-
0018 and ANR-25-CE31-1333 (France);
BMFTR, DFG, HGF, MPG, and AvH Foundation (Germany);
Department of Atomic Energy under Project Identification No.~RTI 4002,
Department of Science and Technology,
and
UPES SEED funding programs
No.~UPES/R\&D-SEED-INFRA/17052023/01 and
No.~UPES/R\&D-SOE/20062022/06 (India);
Israel Science Foundation Grant No.~2476/17,
U.S.-Israel Binational Science Foundation Grant No.~2016113, and
Israel Ministry of Science Grant No.~3-16543;
Istituto Nazionale di Fisica Nucleare and the Research Grants BELLE2,
and
the ICSC – Centro Nazionale di Ricerca in High Performance Computing, Big Data and Quantum Computing, funded by European Union – NextGenerationEU;
Japan Society for the Promotion of Science, Grant-in-Aid for Scientific Research Grants
No.~16H03968,
No.~16H03993,
No.~16H06492,
No.~16K05323,
No.~17H01133,
No.~17H05405,
No.~18K03621,
No.~18H03710,
No.~18H05226,
No.~19H00682, 
No.~20H05850,
No.~20H05858,
No.~22H00144,
No.~22K14056,
No.~22K21347,
No.~23H05433,
No.~26220706,
and
No.~26400255,
and
the Ministry of Education, Culture, Sports, Science, and Technology (MEXT) of Japan;  
National Research Foundation (NRF) of Korea Grants 
No.~2021R1-A6A1A-03043957,
No.~2021R1-F1A-1064008, 
No.~2022R1-A2C-1003993,
No.~2022R1-A2C-1092335,
No.~RS-2016-NR017151,
No.~RS-2018-NR031074,
No.~RS-2021-NR060129,
No.~RS-2023-00208693,
No.~RS-2024-00354342
and
No.~RS-2025-02219521,
Radiation Science Research Institute,
Foreign Large-Size Research Facility Application Supporting project,
the Global Science Experimental Data Hub Center, the Korea Institute of Science and
Technology Information (K25L2M2C3 ) 
and
KREONET/GLORIAD;
Universiti Malaya RU grant, Akademi Sains Malaysia, and Ministry of Education Malaysia;
Frontiers of Science Program Contracts
No.~FOINS-296,
No.~CB-221329,
No.~CB-236394,
No.~CB-254409,
and
No.~CB-180023, and SEP-CINVESTAV Research Grant No.~237 (Mexico);
the Polish Ministry of Science and Higher Education and the National Science Center;
the Ministry of Science and Higher Education of the Russian Federation
and
the HSE University Basic Research Program, Moscow;
University of Tabuk Research Grants
No.~S-0256-1438 and No.~S-0280-1439 (Saudi Arabia), and
Researchers Supporting Project number (RSPD2025R873), King Saud University, Riyadh,
Saudi Arabia;
Slovenian Research Agency and Research Grants
No.~J1-50010
and
No.~P1-0135;
Ikerbasque, Basque Foundation for Science,
State Agency for Research of the Spanish Ministry of Science and Innovation through Grant No. PID2022-136510NB-C33, Spain,
Agencia Estatal de Investigacion, Spain
Grant No.~RYC2020-029875-I
and
Generalitat Valenciana, Spain
Grant No.~CIDEGENT/2018/020;
the Swiss National Science Foundation;
The Knut and Alice Wallenberg Foundation (Sweden), Contracts No.~2021.0174, No.~2021.0299, and No.~2023.0315;
National Science and Technology Council,
and
Ministry of Education (Taiwan);
Thailand Center of Excellence in Physics;
TUBITAK ULAKBIM (Turkey);
National Research Foundation of Ukraine, Project No.~2020.02/0257,
and
Ministry of Education and Science of Ukraine;
the U.S. National Science Foundation and Research Grants
No.~PHY-1913789 
and
No.~PHY-2111604, 
and the U.S. Department of Energy and Research Awards
No.~DE-AC06-76RLO1830, 
No.~DE-SC0007983, 
No.~DE-SC0009824, 
No.~DE-SC0009973, 
No.~DE-SC0010007, 
No.~DE-SC0010073, 
No.~DE-SC0010118, 
No.~DE-SC0010504, 
No.~DE-SC0011784, 
No.~DE-SC0012704, 
No.~DE-SC0019230, 
No.~DE-SC0021616, 
No.~DE-SC0022350, 
No.~DE-SC0023470; 
and
the Vietnam Academy of Science and Technology (VAST) under Grants
No.~NVCC.05.02/25-25
and
No.~DL0000.05/26-27.

These acknowledgements are not to be interpreted as an endorsement of any statement made
by any of our institutes, funding agencies, governments, or their representatives.

We thank G.~Isidori for useful discussions. We thank the SuperKEKB team for delivering high-luminosity collisions;
the KEK cryogenics group for the efficient operation of the detector solenoid magnet and IBBelle on site;
the KEK Computer Research Center for on-site computing support; the NII for SINET6 network support;
and the raw-data centers hosted by BNL, DESY, GridKa, IN2P3, INFN, 
PNNL/EMSL, 
and the University of Victoria.

\end{acknowledgments}

\textit{Data availability} ---
The full Belle and Belle II data are not publicly available. The collaboration will consider requests for access to the data that support this Letter.

\ifthenelse{\boolean{wordcount}}%
\bibliographystyle{apsrev4-2}
\bibliography{references}

\newpage
\section{Supplemental Material}\label{sec:supplementary}

\begin{figure}[!htb]
\centering
\begin{tabular}{c}
 \includegraphics[width=0.8\linewidth]{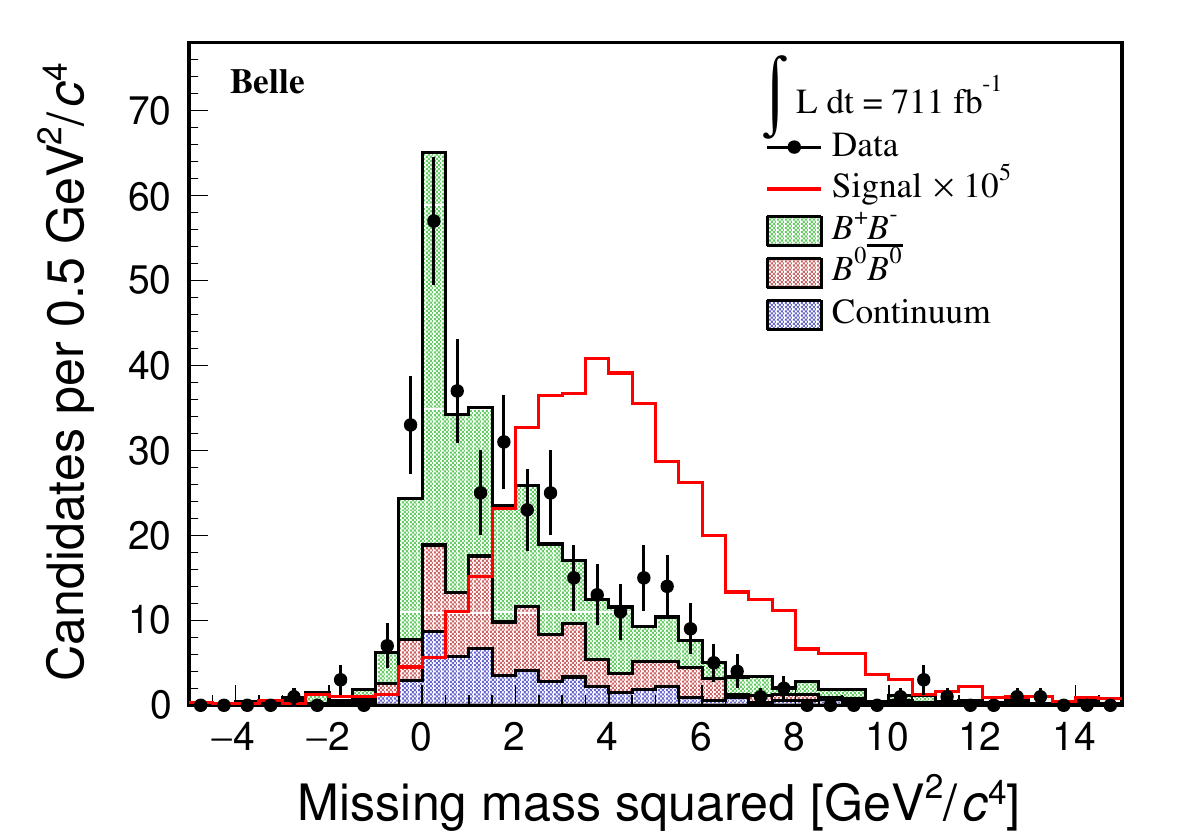}\\
 \includegraphics[width=0.8\linewidth]{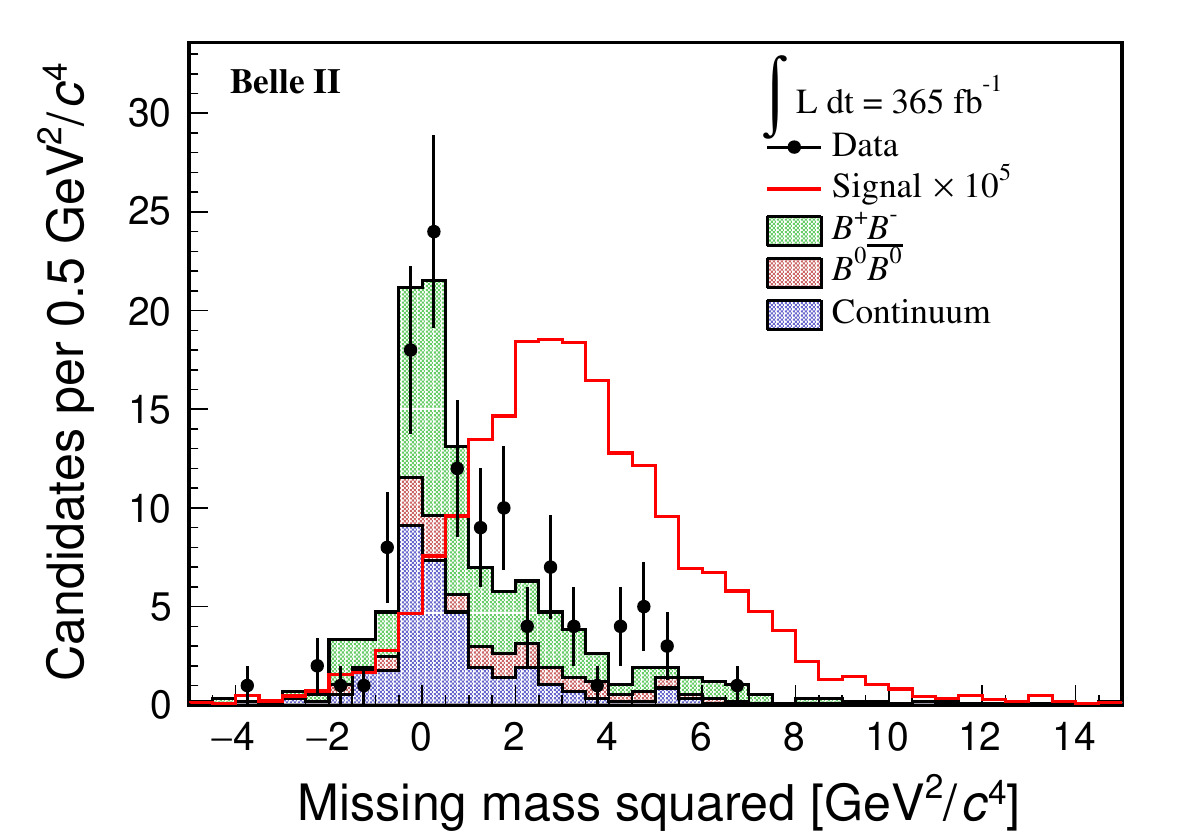}\\
\end{tabular}
\caption{Distribution, with simulation overlaid, of missing mass squared in (top) Belle and (bottom) Belle II events restricted to the $m(K^+\ell^-)> 1.9$ GeV/$c^2$ region.}
 \label{fig:missm2}
\end{figure}

\begin{figure}[!htb]
\centering
\begin{tabular}{c}
 \includegraphics[width=0.8\linewidth]{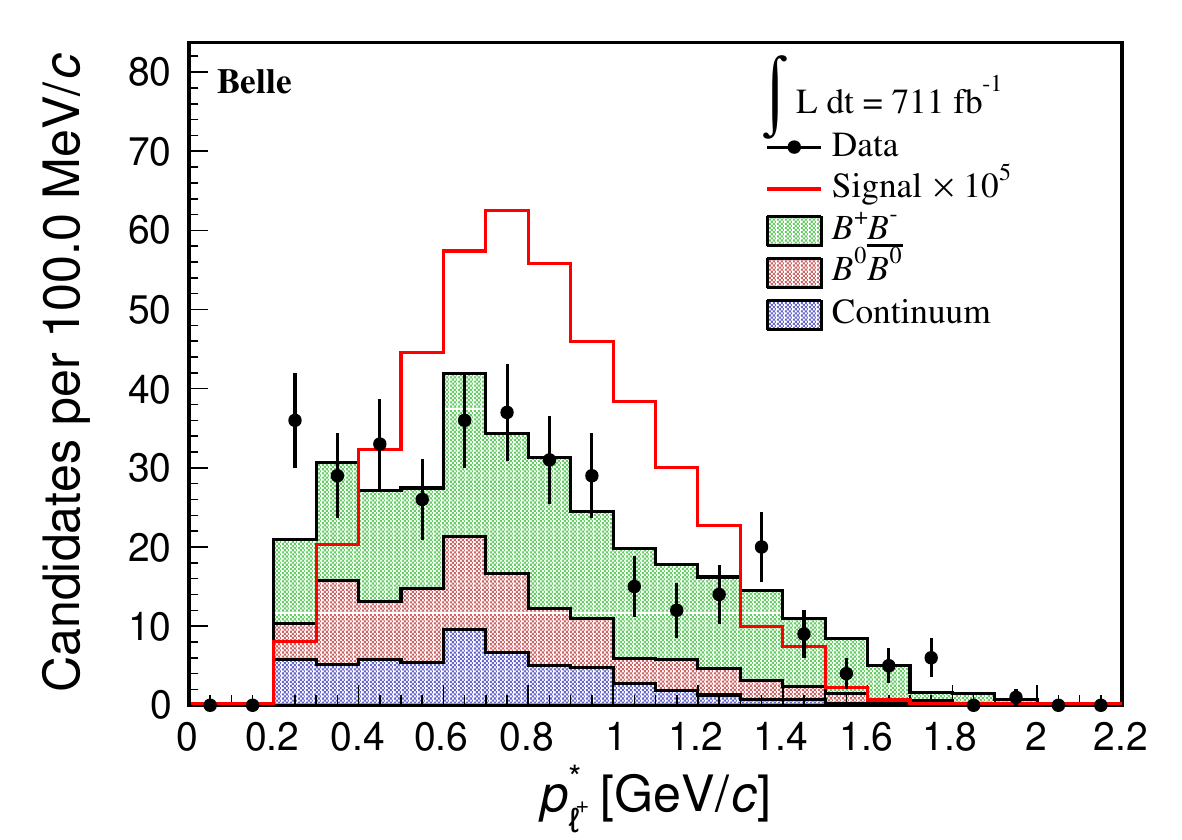}\\
 \includegraphics[width=0.8\linewidth]{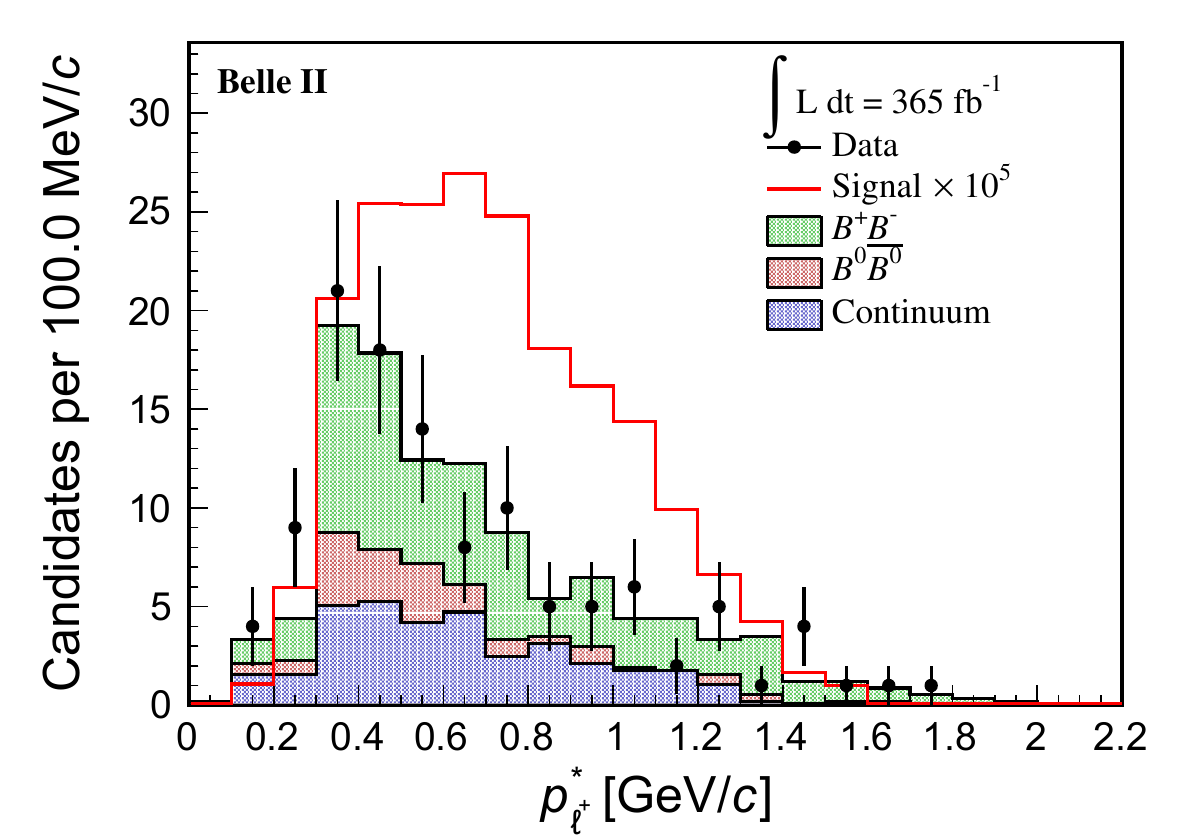}\\
\end{tabular}
\caption{Distribution, with simulation overlaid, of c.m. momentum of the signal lepton with same charge as the signal kaon in (top) Belle and (bottom) Belle II events  restricted to the $m(K^+\ell^-)> 1.9$ GeV/$c^2$ region.}
 \label{fig:plcms}
\end{figure}

\begin{figure}[!htb]
\centering
 \includegraphics[width=0.8\linewidth]{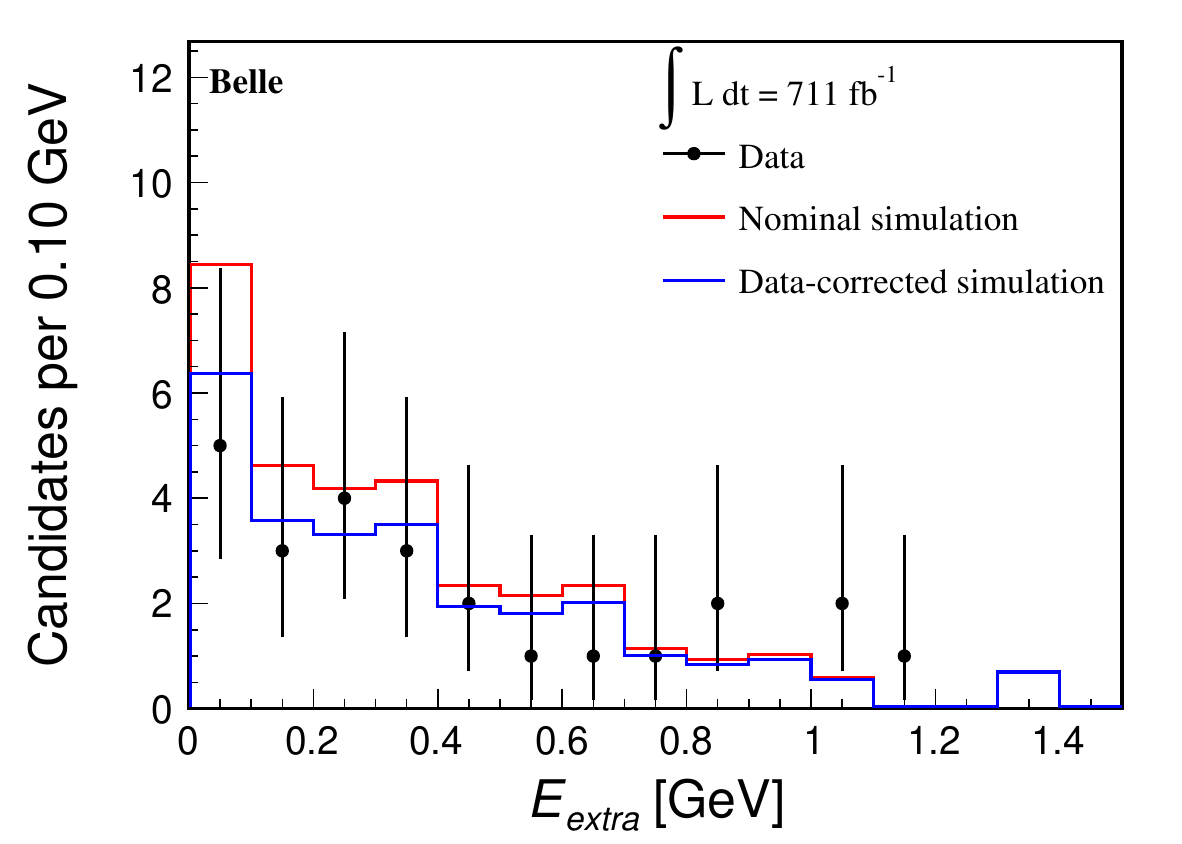} \\
 \includegraphics[width=0.8\linewidth]{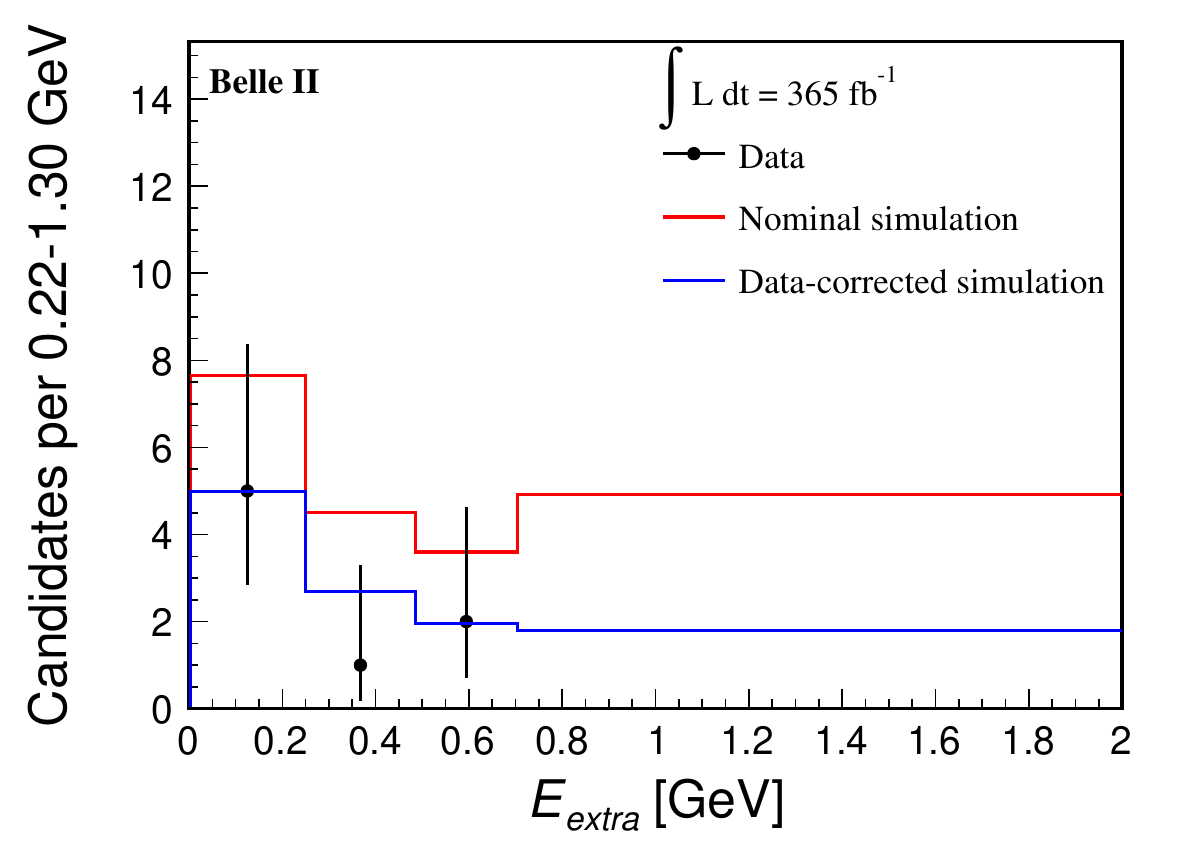} \\
\caption{Distributions of \eecl for (circles) data and (histograms) simulation (red) before and (blue) after the linear-scaling fit for (top) Belle and (bottom) Belle~II events populating the $q^{2}$ sideband.}
 \label{fig:optifit_q2}
\end{figure}

\begin{figure}[!htb]
\centering
 \includegraphics[width=0.8\linewidth]{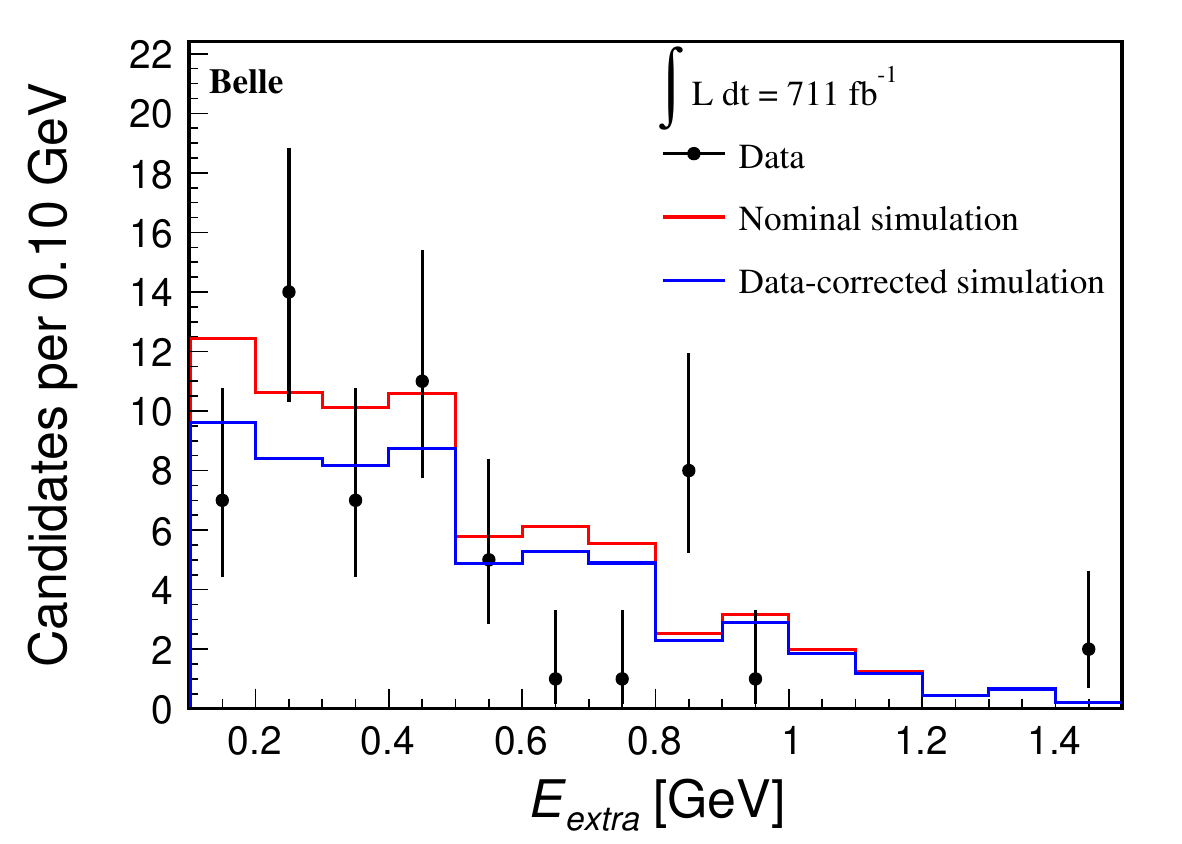}\\
 \includegraphics[width=0.8\linewidth]{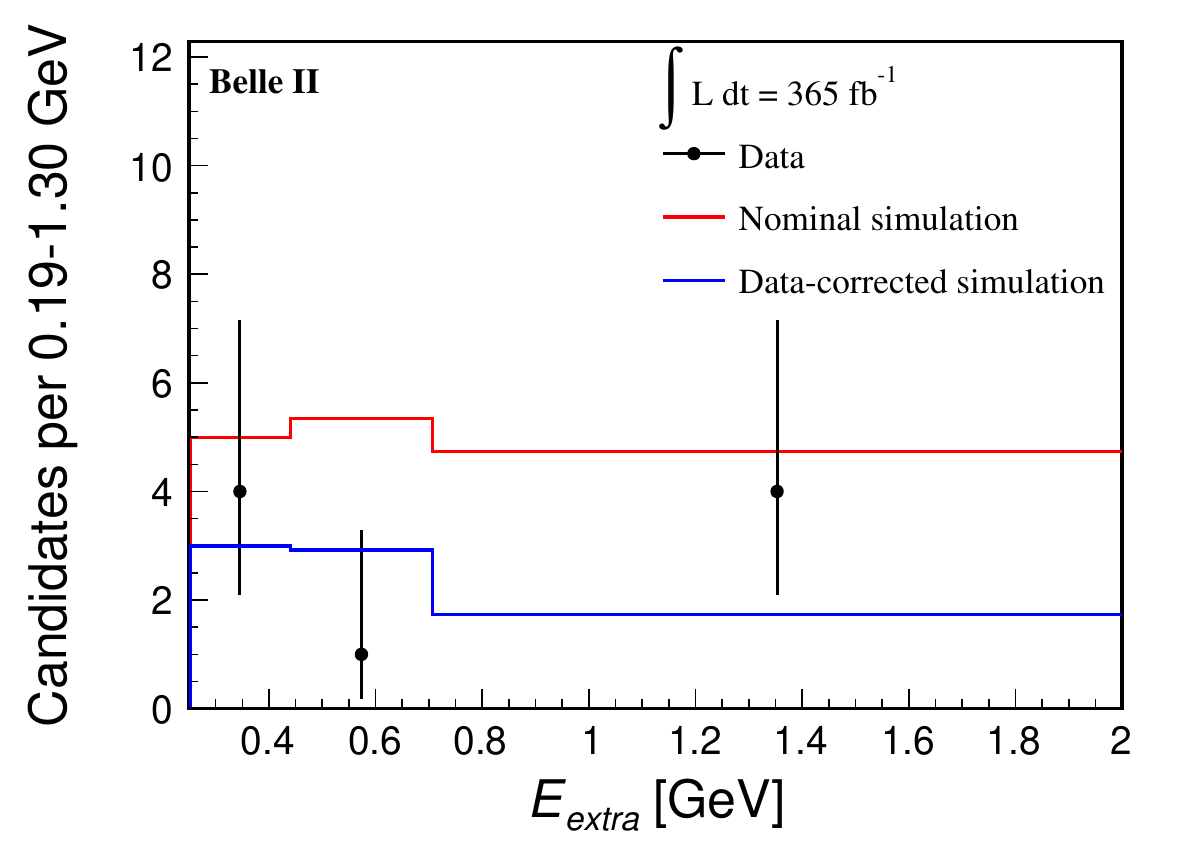} \\
\caption{Distributions of \eecl for (circles) data and (histograms) simulation (red) before and (blue) after the linear-scaling fit for (top) Belle and (bottom) Belle~II events populating the \eecl sideband.}
 \label{fig:optifit_eecl}
\end{figure}

\begin{figure}[!htb]
\centering
\includegraphics[width=0.8\linewidth]{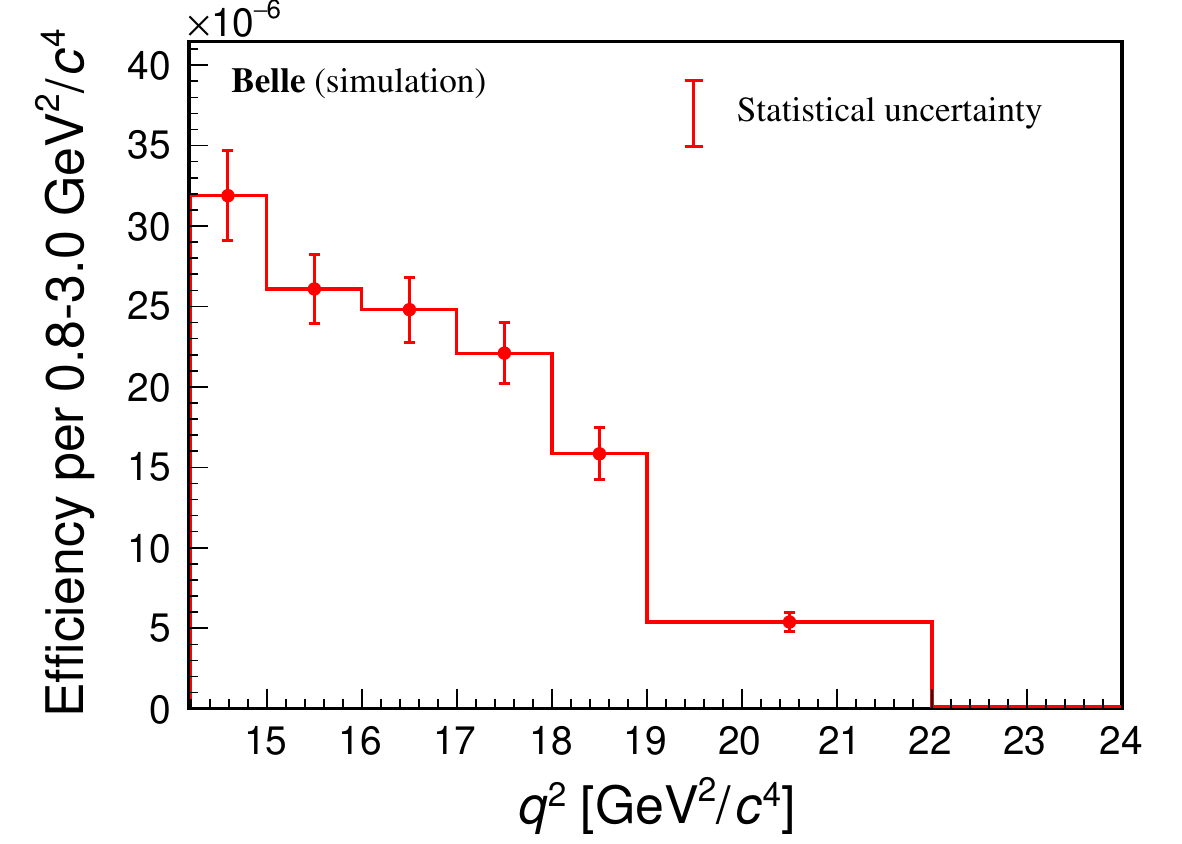}\\
\includegraphics[width=0.8\linewidth]{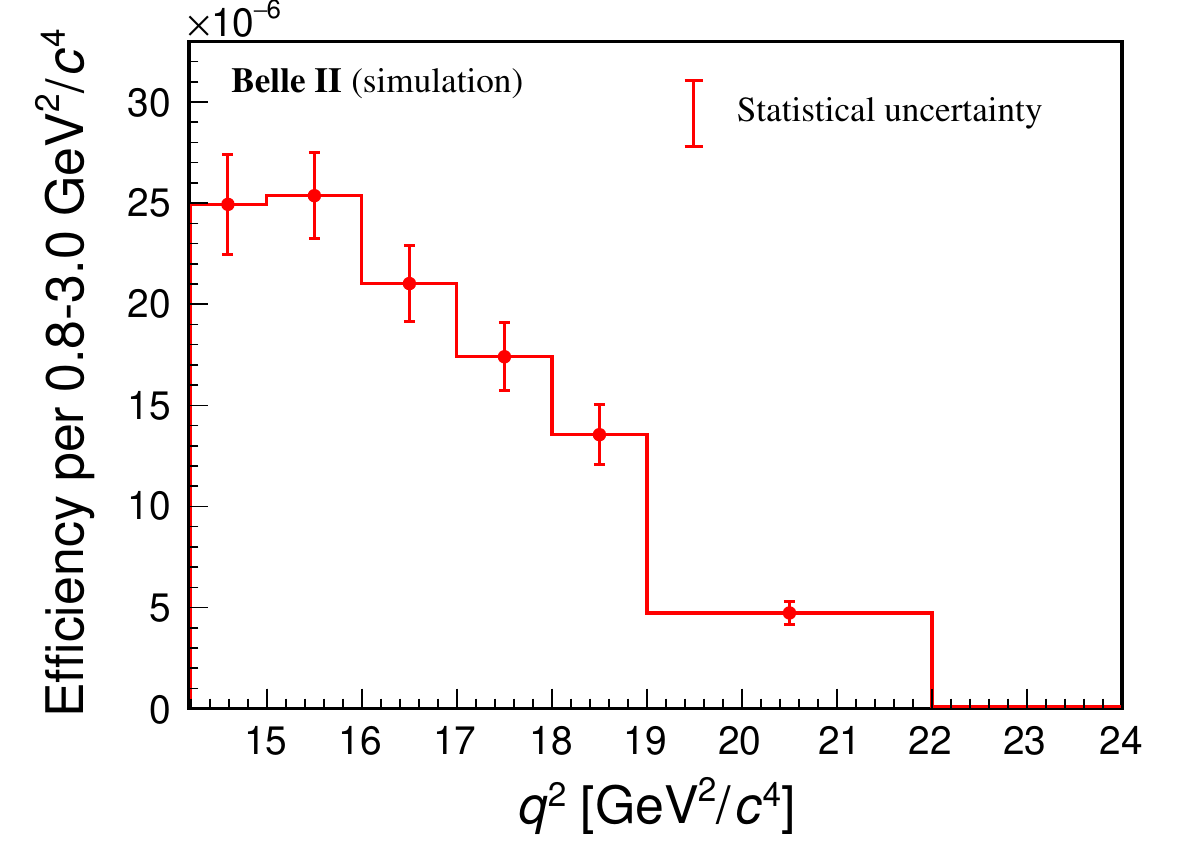} \\
\caption{Distributions of efficiency as a function of $q^{2}$ for reconstructing simulated (top) Belle and (bottom) Belle II signal events. The binning follows Ref.~\cite{Parrott:2022zte}.}
 \label{fig:effvsq2}
\end{figure}

We report here additional figures that help illustrating analysis procedures and their rationales.
Figure~\ref{fig:missm2} shows the distribution of missing mass squared, which is used in the optimized selection. The higher prevalence of neutrinos in the signal decay with respect to the most prominent backgrounds offers distinctive  discrimination. Similarly, Fig.~\ref{fig:plcms} shows the distribution of the c.m.\ momentum of the candidate signal lepton that has the same charge as the candidate signal kaon. Distinctive kinematic-charge correlations produce a higher mean-valued spectrum for signal events compared to backgrounds.
Figures~\ref{fig:optifit_q2} to \ref{fig:optifit_eecl} show the distributions of \eecl before and after the linear-scaling fit that matches simulation to data for events populating the $q^{2}$ and \eecl sidebands. These distributions, along with the analogous distribution of events populating the partner-$B$ sideband reported in the main text body, are simultaneously fit to achieve the final background expectation. Figure~\ref{fig:effvsq2} shows the distributions of simulated signal efficiency as a function of $q^2$ after final selections and corrections.

\end{document}